\documentclass[12pt]{article}
\usepackage{amsmath,amsthm,amssymb}               
\usepackage{graphicx}
\usepackage[left=20mm,right=20mm,top=10mm,bottom=20mm]{geometry}
\usepackage{float}

\def\--{-\kern-.5zw-\kern-.5zw-}

\newcommand{\beq}{\begin{equation}}
\newcommand{\eeq}{\end{equation}}
\newcommand{\bea}{\begin{eqnarray}}
\newcommand{\eea}{\end{eqnarray}}

\newcommand{\dis}{\displaystyle}

\newcommand{\pal}{\partial}

\begin{document}
\setlength{\baselineskip}{18pt}
\begin{titlepage}

\begin{flushright}
KOBE-TH-09-{09}
\end{flushright}
\vspace{1.0cm}
\begin{center}
{\LARGE\bf CP Violation due to Compactification}  
\end{center}
\vspace{25mm}

\begin{center}
{\large
C.S. Lim, 
%
Nobuhito Maru$^*$ 
%
%
and Kenji Nishiwaki
}
\end{center}
\vspace{1cm}
\centerline{{\it Department of Physics, Kobe University,
Kobe 657-8501, Japan}}

\centerline{{\it
$^*$Department of Physics, Chuo University, 
Tokyo 112-8551, Japan
}}
%
%
\vspace{2cm}
\centerline{\large\bf Abstract}
\vspace{0.5cm}
We address the challenging issue of how CP violation is realized in higher dimensional gauge theories without higher dimensional elementary scalar fields. In such theories interactions are basically governed by a gauge principle and therefore to get CP violating phases is a non-trivial task. 
It is demonstrated that  CP violation is achieved as the result of compactification of extra dimensions, which is incompatible with the 
4-dimensional CP transformation. As a simple example we adopt a 6-dimensional U(1) model compactified on a 2-dimensional orbifold $T^{2}/Z_{4}$. We argue that the 4-dimensional CP transformation is related to the complex structure of the extra space and show how the $Z_{4}$ orbifolding leads to CP violation. 
We confirm by explicit calculation of the interaction vertices that CP violating phases remain even after the re-phasing of relevant fields. For completeness, we derive a re-phasing invariant CP violating quantity, following a similar argument in the Kobayashi-Maskawa model which led to the Jarlskog parameter. As an example of a CP violating observable we briefly comment on the electric dipole moment of the electron.    

\end{titlepage} 

\newpage 

\section{Introduction}  

In spite of the great success of the Kobayashi-Maskawa model \cite{KM}, the fundamental origin of
CP violation still seems to be elusive. Once space-time is enlarged such that it contains extra spatial dimensions, some new types of mechanism of CP violation may be possible. 
In this paper we address the question as to whether CP violation is realized as the result of compactification of the extra spatial dimensions. 

In order to extract the new type of CP violating mechanism due to the compactification, we will work in the framework of higher dimensional gauge theories without (higher dimensional) elementary scalar fields. Namely we exclude, e.g. the models of universal extra dimension, where the Higgs scalar is introduced and the same mechanism of CP violation as that in the 
Kobayashi-Maskawa model is operative. 

A typical example of such theories is 10-dimensional supersymmetric Yang-Mills theory, which is the low-energy point particle limit of the open string sector of superstring theory. An interesting and non-trivial question is how to get CP violation in this type of higher dimensional gauge theory. Let us note that in such theories all interactions including possible four-dimensional Yukawa couplings are basically controlled by the gauge principle and therefore, the theory to start with is expected to be CP invariant, since all gauge couplings are of course real. Thus to realize CP violation is a challenging issue. 

Since the original theory is CP invariant, a possible way to break CP would be ``spontaneous violation''. More precisely, one of the few possibilities to break CP symmetry in such theories is to invoke the manner of compactification \cite{Strominger, Lim}, which determines the vacuum state of the theory. (See also ref. \cite{TKobayashi} for a discussion of CP symmetry in orbifold superstring theories.) An important observation 
in the argument is that although C and P transformations in higher dimensions can be easily found such that, 
$\psi^{c} = C \bar{\psi}^{t}, \ \ C^{\dagger} \Gamma^{M} C = - (\Gamma^{{M}})^{t}$ for instance, they do not simply reduce to ordinary 4-dimensional transformations and should be modified in order to recover the 4-dimensional ones. Interestingly, such a modified CP transformation was demonstrated to be equivalent (for even space-time dimensions) to the complex conjugation of the complex homogeneous coordinates $z^{a}$ describing the extra space \cite{Lim}, 
\beq 
CP: \ \ \ z^{a} \ \to \ z^{a \ast}.   
\eeq 
For illustrative purpose, let us consider the four generation model in Type-I superstring theory with six-dimensional Calabi-Yau manifold defined by a quintic polynomial 
\beq 
\sum_{a=1}^{5} (z^{a})^{5} - C (z^{1}z^{2} \cdots z^{5}) = 0.    
\eeq
CP is broken only when the coefficient $C$ is complex, since otherwise the above defining equation is clearly invariant under $z^{a} \to z^{a \ast}$. Another possibility of spontaneous CP violation in this type of theory is 
due to the vacuum expectation value of an effective four-dimensional scalar field, which is originally the extra space component of the gauge field and may have an odd CP eigenvalue \cite{Poland, ALM4}.

Unfortunately, the Calabi-Yau manifold is not easy to handle and to derive the resultant 4-dimensional couplings is very challenging. In this paper, we focus on a much simpler compactification, where interaction vertices are easily obtained. Namely, we discuss CP violation in the six-dimensional U(1) model due to the compactification on a 2-dimensional orbifold $T^{2}/Z_{4}$. We note that the six-dimensional model is the simplest possibility for incorporating a complex structure for the extra-space.  

\begin{figure}[htbp]
 \begin{center}
  \includegraphics[width=50mm]{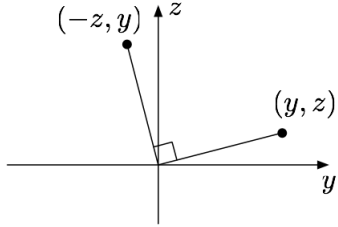}
 \end{center}
 \caption{The identification of points in the orbifold $T^{2}/Z_{4}$.}
 \label{fig:limzu1}
\end{figure}

The $Z_{4}$ orbifolding turns out to lead to CP violation. Without explicit calculations of interaction vertices we can easily understand the reason for CP violation from the following geometrical argument. Let the extra space coordinates be $(y, z)$ and 
combine the pair of coordinates to form a complex coordinate $\omega = \frac{y + iz}{\sqrt{2}}$. The orbifold is obtained by identifying the points related by the action of $Z_4$, the rotation on the $y-z$ plane by an angle $\frac{\pi}{2}$, $(-z, y) \sim (y, z)$ (see Fig.\ref{fig:limzu1}). Or, by use of the complex coordinate, 
\beq 
i \omega \sim \omega. 
\label{orbifoldcondition} 
\eeq 

\begin{figure}[htbp]
 \begin{center}
  \includegraphics[width=50mm]{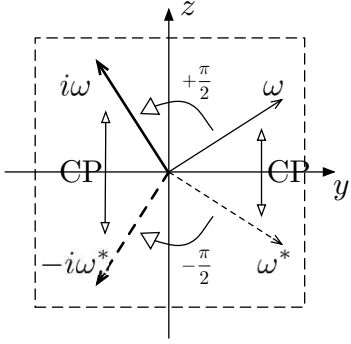}
 \end{center}
 \caption{The CP transformation acting on the orbifold {$T^2/Z_4$}.} 
 \label{fig:orientation-changing}
\end{figure}

As was discussed above and is explicitly shown below (see eq.(\ref{CPincomplexcoord})), in terms of the complex coordinate the CP transformation is known to be equivalent to a complex conjugation \cite{Lim}  
\beq 
CP: \ \ \omega \ \ \to \ \ \omega^{\ast}. 
\eeq
Therefore, as a result of the CP transformation $i \omega$ and $\omega$ in 
 (\ref{orbifoldcondition}) are transformed into $( i \omega)^{\ast}  = -i \omega^{\ast}$ and $\omega^{\ast}$, respectively, and after the CP transformation the orbifold condition becomes
\beq 
(-i) \omega^{{\ast}} \sim \omega^{\ast}.  
\eeq
Namely CP acts as an orientation-changing operator; the rotation by an angle $\frac{\pi}{2}$ has been changed into a rotation by an angle $- \frac{\pi}{2}$. This feature is illustrated in Fig.\ref{fig:orientation-changing}. Hence the orbifold condition is not compatible with the CP transformation and therefore the CP symmetry is broken as the consequence of the orbifold compactification.        
      
\noindent This argument implies that $Z_{2}$ orbifolding does not lead to CP violation, since the identification $- \omega \sim \omega$ is equivalent to $- \omega^{\ast} \sim \omega^{\ast}$, or in other words, a rotation by an angle $\pi$ is equivalent to a rotation by an angle $- \pi$ .  

In the language of the KK(Kaluza-Klein) mode function, CP violation in ${Z_{4}}$ orbifolding may be understood as follows. 
A generic mode function $\phi$ of $(y, z)$ or equivalently of $\omega = {\frac{y+iz}{\sqrt{2}}}$ should have an eigenvalue $t$ under the action of $Z_{4}$, as is seen in eq.(\ref{orbifoldingcond.}) in the next section, 
\beq 
\phi (i \omega) = t \phi (\omega) \ \ \ (t^{4} = 1).  
\label{orbifolding for mode function} 
\eeq
What we obtain from (\ref{orbifolding for mode function}) by taking its complex conjugation is 
\beq 
\phi^{\ast}(-i \omega^{\ast}) = t^{\ast}  \phi^{\ast}(\omega^{\ast}).     
\eeq
This means after the CP transformation, $\phi (\omega) \ \to \ \phi^{\ast}(\omega^{\ast})$, etc., the mode function has eigenvalue $t^{\ast}$ under the action of $Z_{4}$. Thus for mode functions with complex eigenvalues $t = \pm i$ the orbifold condition is not compatible with the CP transformation, leading to the CP violation originating from these mode functions. In fact, we will see in this article that the presence of such mode functions results in CP violating interaction vertices.   

We will discuss how the CP violating phases emerge in the vertices of four-dimensional gauge and Yukawa 
interactions including non-zero KK modes. The phases are confirmed to remain even after the re-phasing of relevant fields by showing a concrete example and also constructing re-phasing invariant quantities. As the typical example of 
CP violating observable, which may be relevant for a model without generation structure, we briefly comment on the electric dipole moment (EDM) of electron.

This paper is organized as follows. 
In the next section, the mode functions on the $T^2/Z_4$ orbifold are constructed and their ortho-normality conditions are derived. In section 3, our model is introduced and after fixing the $Z_{4}$ eigenvalue of each field the fields are expanded as the sum of KK modes by use of the mode functions. In section 4, four-dimensional mass {eigenstates} and corresponding mass eigenvalues are obtained from the free lagrangian, where a $R_{\xi}$-type gauge fixing term for the sector of gauge-Higgs bosons is added. In section 5, we derive gauge and Yukawa interaction vertices with respect to the mass eigenstates. In section 6, we demonstrate, as a simple example, that CP violating phases appear in the interaction vertices of KK photons and argue that the phases remain even after possible re-phasing of the fields. In section 7,  we derive a re-phasing invariant CP violating quantity, following a similar argument in the Kobayashi-Maskawa model \cite{KM}, which led to the Jarlskog parameter \cite{Jar}. In section 8, we briefly comment on the EDM of electron in our model. Section 9 is devoted to the summary.

\section{Mode functions on $T^2/Z_4$}

The ortho-normal set of mode functions on $T^2$ is given as
\beq
\varphi^{(m,n)}(y,z) = \frac{1}{2 \pi R} e^{i\frac{my+nz}{R}} \quad (m,n: \text{integers}).
\eeq
Then the eigenfunctions of $Z_4$ with eigenvalues $t=\pm 1,\pm i$ are constructed by a superposition  
\beq
\tilde{\Phi}^{(m,n)}_t(y,z) = \frac{1}{2} \left[ \varphi^{(m,n)}(y,z) + t^3 \varphi^{(m,n)}(-z,y) +t^2 \varphi^{(m,n)}(-y,-z) +t \varphi^{(m,n)}(z,-y) \right],
\eeq
which satisfies (with $t^4=1$) {\cite{Georgi:2000ks}}
\beq 
\label{orbifoldingcond.}
\tilde{\Phi}^{(m,n)}_t(-z,y) = t \tilde{\Phi}^{(m,n)}_t(y,z){.} 
\eeq

\noindent Namely the eigenfunctions are obtained by the successive action of $Z_4$, the rotation on the $y-z$ plane by an angle $\frac{\pi}{2}$, on the mode functions on $T^2$.

\begin{figure}[H]
 \begin{center}
  \includegraphics[width=50mm]{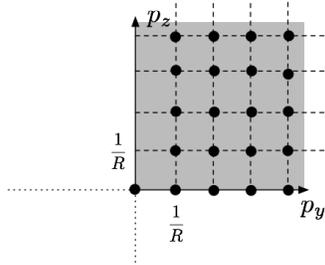}
 \end{center}
\caption{The fundamental domain in the plane of extra space components of momentum.}
 \label{fig:limzu2}
\end{figure}

Let us note $\tilde{\Phi}^{(m,n)}_t$ are also obtained by the rotation in the momentum space $(\frac{m}{R},\frac{n}{R}) \rightarrow (\frac{n}{R},-\frac{m}{R})$:
\beq
\tilde{\Phi}^{(m,n)}_t(y,z) = \frac{1}{2} \left[ \varphi^{(m,n)}(y,z) + t^3 \varphi^{(n,-m)}(y,z) + t^2 \varphi^{(-m,-n)}(y,z) + t \varphi^{(-n,m)}(y,z) \right]. 
\label{momentumexp}
\eeq
This means the extra-dimensional momenta can be restricted to a ``fundamental domain'' shown in Fig. \ref{fig:limzu2} : ($m \geq 1 , n \geq 0$ or $m=n=0$). 
By use of the ortho-normality of $\varphi^{(m,n)}$,
\beq
\int_{-\pi R}^{\pi R}dy \int_{-\pi R}^{\pi R} dz \varphi^{(m,n)}(y,z)^{\ast} \varphi^{(m',n')}(y,z) = \delta_{m,m'}\delta_{n,n'} \quad (m,n,m',n': \text{integers}),
\eeq
and (\ref{momentumexp}), we easily get
\beq
\int_{-\pi R}^{\pi R}dy \int_{-\pi R}^{\pi R} dz \tilde{\Phi}^{(m,n)}_{t} (y,z)^{\ast} \tilde{\Phi}^{(m',n')}_{t'} (y,z) = 
\delta_{m,m'} \delta_{n,n'} \times
\left\{
\begin{array}{ll}
\delta_{t,t'} & (m \geq 1 , n\geq 0) \\
4 \delta_{t,1}\delta_{t',1} & (m=n=0) 
\end{array}
\right. .
\eeq
Thus the orthogonal set of mode functions are known to be ($m \geq 1 , n \geq 0$ or $m=n=0$ (only for $t=1$)) 
\begin{align}
\Phi^{(m,n)}_{t=1}(y,z) &= \frac{1}{\sqrt{1 + 3 \delta_{m,0} \delta_{n,0}}} \tilde{\Phi}^{(m,n)}_{t= 1}(y,z)= \frac{1}{2 \pi R} \frac{1}{\sqrt{1 + 3 \delta_{m,0} \delta_{n,0}}}
\left[ \cos\left( \frac{my+nz}{R} \right) + \cos\left( \frac{ny-mz}{R} \right)  \right], \notag \\
\Phi^{(m,n)}_{t=-1}(y,z) &= \tilde{\Phi}^{(m,n)}_{t=-1}(y,z) = \frac{1}{2 \pi R} \left[ \cos\left( \frac{my+nz}{R} \right) - \cos\left( \frac{ny-mz}{R} \right)  \right], \notag \\
\Phi^{(m,n)}_{t=i}(y,z) &= -i \tilde{\Phi}^{(m,n)}_{t=i}(y,z) = \frac{1}{2 \pi R} \left[ \sin\left( \frac{my+nz}{R} \right) -i \sin\left( \frac{ny-mz}{R} \right)  \right], \notag \\
\Phi^{(m,n)}_{t=-i}(y,z) &= -i \tilde{\Phi}^{(m,n)}_{t=-i}(y,z) = \frac{1}{2 \pi R} \left[ \sin\left( \frac{my+nz}{R} \right) +i \sin\left( \frac{ny-mz}{R} \right)  \right].  
\end{align}
\noindent In terms of these mode functions, a generic bulk field $F(x,y,z)$ is KK mode-expanded as follows depending on it's 
$Z_{4}$-eigenvalue $t$;
\beq
\displaystyle
F(x,y,z) =
\left\{
\begin{array}{ll}
\displaystyle \frac{1}{2 \pi R} F^{(0)}(x) + \frac{1}{2 \pi R} \sum_{m=1}^{\infty} \sum_{n=0}^{\infty} F^{(m,n)}(x) \left[ \cos\left( \frac{my+nz}{R} \right) + \cos\left( \frac{ny-mz}{R} \right)  \right] & (\text{for} \ t=1), \\
\displaystyle \frac{1}{2 \pi R} \sum_{m=1}^{\infty} \sum_{n=0}^{\infty}  F^{(m,n)}(x) \left[ \cos\left( \frac{my+nz}{R} \right) - \cos\left( \frac{ny-mz}{R} \right)  \right] & (\text{for} \ t=-1), \\
\displaystyle \frac{1}{2 \pi R} \sum_{m=1}^{\infty} \sum_{n=0}^{\infty}  F^{(m,n)}(x) \left[ \sin\left( \frac{my+nz}{R} \right) -i \sin\left( \frac{ny-mz}{R} \right)  \right] & (\text{for} \ t=i), \\
\displaystyle \frac{1}{2 \pi R} \sum_{m=1}^{\infty} \sum_{n=0}^{\infty}  F^{(m,n)}(x) \left[ \sin\left( \frac{my+nz}{R} \right) +i \sin\left( \frac{ny-mz}{R} \right)  \right] & (\text{for} \ t=-i).
\end{array}
\right.
\eeq
The presence of the factor $i$ in the fields with $t= \pm i$ signals CP violation. 


\section{The model and  $Z_4$-eigenvalue assignment}

As the simplest realization of CP violation, we consider six-dimensional QED  compactified on $T^2/Z_4$,
whose lagrangian is given as
\beq
\mathcal{L}_{QED} = \overline{\Psi_6} \left\{ \Gamma^M (i \pal_M + gA_M) - m_B \right\} \Psi_6
- \frac{1}{4} (\pal_M A_N - \pal_N A_M) (\pal^M A^N - \pal^N A^M) \ (M,N=0-3, y, z),
\label{eq:lagrangian}
\eeq
\noindent
where gauge-fixing and F-P ghost terms have not been shown explicitly. Let us note that in contrast to the case of a five-dimensional model with $S^1/Z_2$ orbifold, the bulk mass term $- m_B \overline{\Psi_6} \Psi_6$ is allowed,
since $Z_4$ is a rotation in the $y-z$ plane, under which $\overline{\Psi_6} \Psi_6$ is obviously invariant.
The electron described by the zero-mode of (the half of) $\Psi_6$ thus has a mass $m_B$. 
Note also that $A_y$ and $A_z$ have non-trivial $Z_4$-eigenvalues, as is discussed later, and therefore, have neither zero-modes
nor VEV's.

The $Z_4$ symmetry implies that the extra-space components of $A_M$, i.e. $A_y$ and $A_z$, and $\Psi_6$
should properly transform under the action of $Z_4$. First, defining a complexified coordinate and vector potential
\beq
\omega \equiv \frac{y+iz}{\sqrt{2}}, \quad A_{\omega} \equiv \frac{A_y -i A_z} {\sqrt{2}} ,
\eeq
the transformation properties of $A_y$ and $A_z$ are equivalent to
\beq
A_{\omega} (x,i \omega) = (-i) A_{\omega}(x,\omega).
\eeq
Namely, $A_{\omega}$ is an eigenfunction under $Z_4$ with eigenvalue $-i$ and is mode-expanded as
\beq
A_{\omega} (x,y,z) = \frac{1}{2 \pi R} \sum_{m=1}^{\infty} \sum_{n=0}^{\infty} A_{\omega}^{(m,n)}(x)
\left[ \sin\left(\frac{my+nz}{R}\right) +i \sin\left(\frac{ny-mz}{R}\right) \right],
\label{eq:modeexp_of_-i}
\eeq
where $A^{(m,n)}_{\omega}$ are complex functions, whose real and imaginary parts are denoted by $A_y^{(m,n)}$ and $A_z^{(m,n)}$, respectively: $A_{\omega}^{(m,n)} \equiv \frac{A_y^{(m,n)} -i A_z^{(m,n)}}{\sqrt{2}}$.   

Since $Z_4$ is a rotation of an angle $\frac{\pi}{2}$, the 6-dimensional Dirac fermion transforms as
\beq
\frac{\mathbf{I} - \Gamma^y \Gamma^z}{\sqrt{2}} \Psi_6(x,i \omega) = (-i)^{\frac{1}{2}} \Psi_6(x,\omega),
\label{eq:spinor-rotation}
\eeq
where the phase-factor has an arbitrariness and is chosen such that $\Psi_6$ has a zero-mode.
We decompose $\Psi_6$ into two four-dimensional Dirac spinors:
\beq
\Psi_6 \equiv \begin{pmatrix} \psi \\ \Psi \end{pmatrix}.
\eeq
In this base,
\beq
\Gamma^{\mu} = \gamma^{\mu} \otimes I_2 =  \begin{bmatrix} \gamma^{\mu} & \\ & \gamma^{\mu} \end{bmatrix} \ , \ 
\Gamma^y = \gamma^5 \otimes i \sigma_1 = \begin{bmatrix} & i \gamma^5 \\ i \gamma^5 & \end{bmatrix}\ ,\ 
\Gamma^z = \gamma^5 \otimes i \sigma_2 = \begin{bmatrix} & \gamma^5 \\ - \gamma^5 & \end{bmatrix}.
\label{eq:gammas}
\eeq
Then from (\ref{eq:spinor-rotation}) and (\ref{eq:gammas}) we find
\beq
\psi(x,i \omega) = (-i) \psi(x,\omega), \quad \Psi(x,i \omega) = \Psi(x,\omega). 
\eeq
Let us note that only $\Psi$ is allowed to have a zero-mode. Accordingly, each field is mode-expanded as  
\begin{align}
\psi(x,y,z) &= \frac{1}{2 \pi R} \sum_{m=1}^{\infty} \sum_{n=0}^{\infty} \psi^{(m,n)}(x)
\left[ \sin\left(\frac{my+nz}{R}\right) +i \sin\left( \frac{ny-mz}{R} \right) \right], \notag \\
\Psi(x,y,z) &= \frac{1}{2 \pi R} \Psi^{(0)}(x) +  \frac{1}{2 \pi R} \sum_{m=1}^{\infty} \sum_{n=0}^{\infty} \Psi^{(m,n)}(x)
\left[ \cos\left(\frac{my+nz}{R}\right) + \cos\left( \frac{ny-mz}{R} \right) \right].
\label{eq:modeexp_of_fermions}
\end{align}
We will show below that the factor $i$ in front of $\sin\left( \frac{ny-mz}{R} \right)$ in the mode expansion of $\psi (x,y,z)$ results in CP violating phases in the interaction vertices of $A_{\mu}$ with $\psi$.

One may wonder if the requirement of anomaly cancellation affects the CP violation, by enforcing the introduction of additional fields. Fortunately, our model is easily shown to be free from both four-dimensional and six-dimensional anomalies, and there is no need for additional fields. First, the four-dimensional anomaly due to the zero-mode $\Psi^{(0)}(x)$ trivially vanishes, since $\Psi^{(0)}(x)$ is a four-dimensional Dirac spinor and its coupling to the photon is vector-like. Concerning the six 
dimensional anomaly \cite{Dobrescu}, we note that each of $\psi$ and $\Psi$ is ``non-chiral'' in a six-dimensional sense. Namely, each fermion has both eigenvalues $\pm 1$ of 
$\Gamma_{7} \equiv \Gamma^{0} \Gamma^{1} \cdots \Gamma^{y} \Gamma^{z}$. 
This is easily seen, since in the base (\ref{eq:gammas}) $\Gamma_{7} = \gamma^{5} \otimes \sigma_3$ and the eigenvalue of $\Gamma_{7}$ is the product of  the eigenvalues of the four-dimensional chirality and extra-dimensional chirality, namely the eigenvalues of $\gamma^{5}$ and $\sigma_{3}$, respectively. Each of $\psi$ and $\Psi$ has eigenvalues $1$ and $-1$ of $\sigma_{3}$, respectively, while each spinor  is a four-dimensional Dirac spinor and has both eigenvalues, $\pm 1$ of 
$\gamma^{5}$. Thus, each of $\psi$ and $\Psi$ has both eigenvalues $\pm 1$ of 
$\Gamma_{7}$. These properties come essentially from the fact that we have started with a 6-dimensional Dirac spinor. Hence, each of $\psi$ and $\Psi$, being  ``non-chiral'' in six-dimensional sense, does not yield any six-dimensional anomalies.  

In the base of gamma matrix (\ref{eq:gammas}) the ``modified'' P and C transformations, corresponding to ordinary four-dimensional ones, are explicitly given as 
\beq
P: \ \ \Psi_6 \ \ \to \ \  (\gamma^{0} \otimes \sigma_{3}) \Psi_6, \ \ \ 
C: \ \ \Psi_6 \ \ \to \ \  (c_{4} \otimes {\sigma_{3}}) \bar{\Psi}_6^{t} \ \ (c_{4} = i { \gamma^{2}\gamma^{0}}).     
\label{CPtransf.}
\eeq
We can easily check that under the $C$ transformation defined by (\ref{CPtransf.}), a pair of bi-linears of $\Psi_6$, namely $(V^{y}, V^{z}) = ( \bar{\Psi}_{6} \Gamma^{y} \Psi_{6}, \bar{\Psi}_{6} \Gamma^{z} \Psi_{6})$ transforms into $(- V^{y}, V^{z})$, while  $(V^{y}, V^{z})$ is invariant under the $P$ transformation. Accordingly, a pair of extra space coordinates $(y, z)$ should transform as 
\bea  
P: \ \ \ (y, z) \ \ &\to& \ \ (y, z),  \nonumber \\ 
C, \ CP: \ \ \ (y, z) \ \ &\to& \ \ (y, - z).    
\label{CandCP}
\eea
We thus explicitly confirm the transformation property of the complex coordinate discussed in \cite{Lim}:
\beq  
CP: \ \ \ \omega \to \omega^{\ast}. 
\label{CPincomplexcoord} 
\eeq  
 

\section{The mass eigenstates and mass eigenvalues}

Substituting the mode expansions (\ref{eq:modeexp_of_-i}) and (\ref{eq:modeexp_of_fermions}),
together with
\beq
A_{\mu}(x,y,z) = \frac{1}{2 \pi R} A_{\mu}^{(0)}(x) + \frac{1}{2 \pi R} \sum_{m=1}^{\infty} \sum_{n=0}^{\infty}
A_{\mu}^{(m,n)}(x) \left[ \cos\left( \frac{my+nz}{R} \right) + \cos\left( \frac{ny-mz}{R} \right)  \right], 
\label{eq:KK-expansion-of-Amu}
\eeq
in the lagrangian (\ref{eq:lagrangian}) and integrating over the extra-space coordinates $y$ and $z$ we
get the effective theory from four-dimensional perspective. 

We first focus on the free lagrangian to get the mass matrices for various fields of a fixed KK mode.
Let us note that there should be a mixing between $\psi$ and $\Psi$ for fermions and a mixing between $A_{\mu}$
and a certain linear combination of $A_y$ and $A_z$ through a Higgs-like mechanism,
both only for non-zero KK modes.
To get mass eigenstates and their mass eigenvalues, we need to diagonalize the mass matrix in the base of
$\psi$ and $\Psi$ for fermions and put a suitable gauge fixing term to eliminate mixing for the gauge-Higgs sector.

The mass matrix for the fermion in the base of $\psi^{(m,n)}$ and $\Psi^{(m,n)}$ for a given non-zero mode $(m,n)\ (m \geq 1,n \geq 0)$ can be read off from the part $\overline{\Psi_6} \left\{ i(\Gamma^y \pal_y + \Gamma^z \pal_z) - m_B \right\} \Psi_6$. 
After the $y, z$ integrations this part yields the mass term 
\beq
\left( \frac{m-in}{R} \overline{\psi^{(m,n)}} \gamma^5 \Psi^{(m,n)} + h.c. \right) - m_B \left( \overline{\psi^{(m,n)}} \psi^{(m,n)} + \overline{\Psi^{(m,n)}} \Psi^{(m,n)} \right).
\eeq
In order to eliminate $\gamma^5$ in the first parenthesis, we perform a chiral rotation,
\beq
\Psi^{(m,n)} \rightarrow \tilde{\Psi}^{(m,n)} \equiv \gamma^5 \Psi^{(m,n)}.
\eeq
Then in terms of $\psi^{(m,n)}$ and $\tilde{\Psi}^{(m,n)}$ the mass term is written as
\beq
\left( \frac{m-in}{R} \overline{\psi^{(m,n)}}  \tilde{\Psi}^{(m,n)} + h.c. \right) - m_B \overline{\psi^{(m,n)}} \psi^{(m,n)} 
+ m_B \overline{\tilde{\Psi}^{(m,n)}} \tilde{\Psi}^{(m,n)},
\eeq
whose mass matrix is now hermitian, i.e.
\beq
M_f^{(m,n)} =
\begin{pmatrix}
m_B && - \frac{m-in}{R} \\
- \frac{m+in}{R} && - m_B
\end{pmatrix}.
\eeq
The matrix $M_f^{(m,n)}$ is diagonalized by an unitary matrix $U^{(m,n)}$,
\begin{align}
U^{(m,n)\dagger} M_f^{(m,n)} U^{(m,n)} &=
\begin{pmatrix}
m_{f}^{(m,n)} && 0 \\
0 && - m_{f}^{(m,n)}
\end{pmatrix}, \notag \\
U^{(m,n)} &= 
\begin{pmatrix}
1 && 0 \\
0 && e^{i \varphi^{(m,n)}}
\end{pmatrix}
\begin{pmatrix}
\cos{\theta^{(m,n)}} &&  \sin{\theta^{(m,n)}} \\
- \sin{\theta^{(m,n)}} && \cos{\theta^{(m,n)}}
\end{pmatrix} \notag \\
&=
\begin{pmatrix}
\cos{\theta^{(m,n)}} && \sin{\theta^{(m,n)}} \\
- \sin{\theta^{(m,n)}} e^{i \varphi^{(m,n)}} && \cos{\theta^{(m,n)}} e^{i \varphi^{(m,n)}}
\end{pmatrix} 
\label{fermiondiag}
\end{align}
where
\beq
m_{f}^{(m,n)} \equiv \sqrt{m_B^2 + \frac{m^2 + n^2}{R^2}} \ , \ \tan{\varphi^{(m,n)}} \equiv \frac{n}{m} \ ,\ \tan{2 \theta^{(m,n)}} \equiv \frac{\frac{\sqrt{m^2+n^2}}{R}}{m_B}.
\eeq
Let us note that by a chiral transformation to change the sign of the eigenvalue $- m_{f}^{(m,n)}$ we have degenerate mass eigenvalues and then a further unitary transformation by an arbitrary unitary matrix $V^{(m,n)}$ becomes possible.
Thus, the mass eigenstates $\psi^{'(m,n)}$ and $\Psi^{'(m,n)}$ are related to the original fields as
\beq
\begin{pmatrix}
\psi^{(m,n)} \\ \tilde{\Psi}^{(m,n)}
\end{pmatrix}
=
U^{(m,n)} \begin{pmatrix} 1 && 0 \\ 0 && \gamma_5 \end{pmatrix} V^{(m,n)}
\begin{pmatrix}
\psi^{'(m,n)} \\ \Psi^{'(m,n)}
\end{pmatrix},
\eeq 
or in terms of Weyl fermions \footnote{{We define four-dimensional Weyl fermions as $\psi_{R,L} \equiv \frac{1 \pm \gamma^5}{2} \psi.$}} as
\begin{align}
\begin{pmatrix}
\psi^{(m,n)}_R \\ \tilde{\Psi}^{(m,n)}_R
\end{pmatrix}
&= U^{(m,n)} V^{(m,n)}
\begin{pmatrix}
\psi^{'(m,n)}_R \\ \Psi^{'(m,n)}_R
\end{pmatrix}, \notag \\
\begin{pmatrix}
\psi^{(m,n)}_L \\ \tilde{\Psi}^{(m,n)}_L
\end{pmatrix}
&= U^{(m,n)} \begin{pmatrix} 1 && 0 \\ 0 && - 1 \end{pmatrix} V^{(m,n)}
\begin{pmatrix}
\psi^{'(m,n)}_L \\ \Psi^{'(m,n)}_L
\end{pmatrix}.
\label{eq:left-rotation}
\end{align}
The freedom of $V^{(m,n)}$ may signal the internal symmetry between two Dirac fermions obtained from a massive 
six-dimensional Dirac fermion by dimensional reduction. Any physical observables should be invariant under the unitary transformation due to $V^{(m,n)}$. So, without loss of generality we can choose a base 
where $V^{(m,n)} = U^{(m,n)\dagger}$, to get
\begin{align}
\begin{pmatrix}
\psi_R^{(m,n)} \\ \tilde{\Psi}_R^{(m,n)}
\end{pmatrix}
&=
\begin{pmatrix}
\psi_R^{'(m,n)} \\ \Psi_R^{'(m,n)}
\end{pmatrix}, \notag \\
\begin{pmatrix}
\psi_L^{(m,n)} \\ \tilde{\Psi}_L^{(m,n)}
\end{pmatrix}
&= \hat{U}^{(m,n)}
\begin{pmatrix}
\psi_L^{'(m,n)} \\ \Psi_L^{'(m,n)}
\end{pmatrix},
\label{eq:u-hat-basis}
\end{align}  
where the unitary and hermitian matrix $\hat{U}^{(m,n)} \ (\hat{U}^{(m,n)\dagger} = \hat{U}^{(m,n)}, \ (\hat{U}^{(m,n)})^{2} = I_{2})$ is given as (as is easily derived from (\ref{fermiondiag}))
\begin{align}
\hat{U}^{(m,n)} &= U^{(m,n)} \begin{pmatrix} 1 & 0 \\ 0 & - 1 \end{pmatrix} U^{(m,n)\dagger}
= \frac{1}{m_{f}^{(m,n)}} M_{f}^{(m,n)} 
 \notag \\
&= \frac{1}{m_{f}^{(m,n)}} 
\begin{pmatrix}
m_B & - \frac{m-in}{R} \\ - \frac{m+in}{R} & -m_B
\end{pmatrix}.
\label{eq:u-matrix}
\end{align}
$\hat{U}^{(m,n)}$ denotes the asymmetry in the unitary transformations (\ref{eq:left-rotation}) between the right-handed and
left-handed fermions, which is of real physical interest.
The mass term for non-zero KK modes is thus written with degenerate mass $m^{(m,n)}_{f}$ as
\beq
- m_{f}^{(m,n)} \left( \overline{\psi^{'(m,n)}} \psi^{'(m,n)} + \overline{\Psi^{'(m,n)}} \Psi^{'(m,n)} \right),
\label{eq:physicalmassterm}
\eeq
where
\beq
\psi^{'(m,n)} \equiv \psi^{'(m,n)}_R + \psi^{'(m,n)}_L\ ,\ \Psi^{'(m,n)} \equiv \Psi^{'(m,n)}_R + \Psi^{'(m,n)}_L.
\label{eq:4D-Diracfermion}
\eeq

Concerning the mass term for the zero mode, there is no mixing between $\psi$ and $\Psi$, since only the state $\Psi$ exists:  
$\Psi^{'(0)} = \Psi^{(0)}$. The mass term takes a simple form
\beq
- m_B \overline{\Psi^{'(0)}} \Psi^{'(0)}.
\label{eq:zeromode-massterm}
\eeq
Let us note that $\Psi^{'(0)} = \Psi^{(0)}$ and (\ref{eq:zeromode-massterm}) are naturally obtained formally 
setting $m=n=0$ in (\ref{eq:u-hat-basis}) , (\ref{eq:u-matrix}) and (\ref{eq:physicalmassterm}).

We now move to the part relevant for the mass-squared of gauge-Higgs bosons:
\beq
- \frac{1}{2} \left( - F_{\mu y} F^{\mu}_{\ \ y} - F_{\mu z} F^{\mu}_{\ \ z} + F_{yz}^2 \right) \quad (F_{\mu y} = \pal_{\mu} A_y - \pal_y A_{\mu}\ ,\ \text{etc.}),
\eeq
which may be written in terms of $A_{\omega}$ as,
\begin{align}
& (\pal_{\mu} A_{\omega})(\pal^{\mu} A_{\overline{\omega}}) + (\pal_{\omega} A_{\mu})(\pal_{\overline{\omega}} A^{\mu})
- \left\{ (\pal_{\mu} A_{\omega})(\pal_{\overline{\omega}} A^{\mu}) + (\pal_{\mu} A_{\overline{\omega}})(\pal_{\omega} A^{\mu}) \right\} \notag \\
& + \frac{1}{2} (\pal_{\overline{\omega}} A_{\omega} - \pal_{\omega} A_{\overline{\omega}})^2
\quad \left(A_{\overline{\omega}} = A_{\omega}^{\ast}\ ,\ \pal_{\omega} = \frac{\pal_y -i \pal_z}{\sqrt{2}}\ ,\ \text{etc.} \right).
\label{eq:omega-massterm}
\end{align}
Since the sector of non-zero KK modes possess a Higgs-like mechanism, in order to form four-dimensional massive gauge bosons,
we now introduce the gauge fixing term $\grave{a}$ la $R_{\xi}$ gauge \cite{FLS}:
\beq
- \frac{1}{2 \xi} \left\{ \pal_{\mu} A^{\mu} - \xi \left( \pal_{\omega} A_{\overline{\omega}} + \pal_{\overline{\omega}}
A_{\omega} \right) \right\}^2.
\label{eq:gauge-fixing-term}
\eeq  
The aim is to eliminate the term in (\ref{eq:omega-massterm}),
which may be rewritten, after partial integrals, as $- (\pal_{\omega} A_{\overline{\omega}} + \pal_{\overline{\omega}} A_{\omega}) \pal_{\mu} A^{\mu}$.

Combining (\ref{eq:omega-massterm}) and (\ref{eq:gauge-fixing-term}) we get
\begin{align}
& (\pal_{\omega} A_{\mu})(\pal_{\overline{\omega}} A^{\mu}) - \frac{1}{2 \xi} (\pal_{\mu} A^{\mu})^2 \notag \\
&+(\pal_{\mu} A_{\omega})(\pal^{\mu} A_{\overline{\omega}}) -\frac{\xi}{2} (\pal_{\omega} A_{\overline{\omega}} + \pal_{\overline{\omega}} A_{\omega})^2 + \frac{1}{2} (\pal_{\overline{\omega}} A_{\omega} - \pal_{\omega} A_{\overline{\omega}})^2.
\label{eq:total-massterm}
\end{align}
We thus realize that $\text{Re}(\pal_{\overline{\omega}} A_{\omega})$ and $\text{Im}(\pal_{\overline{\omega}} A_{\omega})$,
i.e.
$\frac{1}{\sqrt{m^2 + n^2}} (m A_y^{(m,n)} + n A_z^{(m,n)})$ and $\frac{1}{\sqrt{m^2 + n^2}} (-n A_y^{(m,n)} + m A_z^{(m,n)})$ 
 behave as would-be N-G boson and physical Higgs boson,
respectively, for non-zero KK modes. 

In fact, substituting (\ref{eq:modeexp_of_-i}) and (\ref{eq:KK-expansion-of-Amu}) in (\ref{eq:total-massterm}) and
integrating over $y$ and $z$, we get corresponding four-dimensional effective lagrangian,
\begin{align}
& \frac{1}{2} \sum_{m=1}^{\infty} \sum_{n=0}^{\infty} \frac{m^2 + n^2}{R^2} A_{\mu}^{(m,n)} A^{\mu (m,n)}
- \frac{1}{2 \xi} \left\{ (\pal_{\mu} A^{(0)\mu})^2 + \sum_{m=1}^{\infty} \sum_{n=0}^{\infty} (\pal_{\mu} A^{(m,n) \mu})^2
\right\} \notag \\
&+\frac{1}{2} \sum_{m=1}^{\infty} \sum_{n=0}^{\infty} \left\{ (\pal_{\mu} G^{(m,n)})(\pal^{\mu} G^{(m,n)}) + (\pal_{\mu} H^{(m,n)})(\pal^{\mu} H^{(m,n)}) - \xi \frac{m^2 + n^2}{R^2} {G^{(m,n)}}^2 -  \frac{m^2 + n^2}{R^2} {H^{(m,n)}}^2 \right\} ,
\end{align}
where $G^{(m,n)}$ and $H^{(m,n)}$ denote the would-be N-G boson and the physical Higgs boson :
\beq
\left\{
\begin{array}{l}
\dis G^{(m,n)}(x) = \frac{1}{\sqrt{m^2 +n^2}} \left( m A_y^{(m,n)}(x) + n A_z^{(m,n)}(x) \right), \\
\dis H^{(m,n)}(x) = \frac{1}{\sqrt{m^2 +n^2}} \left( -n A_y^{(m,n)}(x) + m A_z^{(m,n)}(x) \right).
\end{array}
\right.
\eeq
It is now clear that we get a massless photon $A_{\mu}^{(0)}$, along with a massive photon $A_{\mu}^{(m,n)}$
and massive Higgs boson $H^{(m,n)}$, both having masses $M_{V}^{(m,n)} \equiv \frac{\sqrt{m^2 + n^2}}{R}$, in the four-dimensional spectrum.


\section{The interaction vertices}

Having KK-mode expansions for each field, we are now ready to calculate the interaction vertices in terms of four-dimensional
fields.
First, we focus on the interaction vertices of four-dimensional gauge fields $A_{\mu}^{(m,n)}$ (and $A_{\mu}^{(0)}$).
Since the interaction preserves the chirality of fermions and the right-handed fermions are not associated
with unitary transformation when described by mass eigenstates (see (\ref{eq:u-hat-basis}))
we initially restrict ourselves to the interaction vertices for the right-handed fermions.
The relevant part of the lagrangian is ($\tilde{\Psi} \equiv \gamma^5 \Psi$)
\beq
g \left( \overline{\psi} \gamma^{\mu} R \psi + \overline{\tilde{\Psi}} \gamma^{\mu} R \tilde{\Psi} \right) A_{\mu}.
\label{eq:gaune-interactions}
\eeq
Substituting (\ref{eq:modeexp_of_fermions}) and (\ref{eq:KK-expansion-of-Amu}) in (\ref{eq:gaune-interactions}), with the mode expansion for $\Psi$ being modified as
\beq
\tilde{\Psi}(x,y,z) = \frac{1}{2 \pi R} \gamma^5 \Psi^{(0)}(x) + \frac{1}{2 \pi R} \sum_{m=1}^{\infty} \sum_{n=0}^{\infty}
\tilde{\Psi}^{(m,n)}(x) \left[ \cos\left( \frac{my+nz}{R} \right) + \cos\left( \frac{mz-ny}{R} \right) \right],
\eeq 
we get after $y$ and $z$ integrations the interaction vertices with respect to non-zero KK modes, 
\begin{align}
{\sum_{m,m',m''=1}^{\infty} \sum_{n,n',n''=0}^{\infty}}
\frac{g_4}{2}
\begin{pmatrix} \overline{\psi^{'(m,n)}}(x) & \overline{\Psi^{'(m,n)}}(x) \end{pmatrix}
U_{R(m,n;m'',n'')}^{(m',n')} \gamma^{\mu} R
\begin{pmatrix} \psi^{'(m'',n'')}(x) \\ \Psi^{'(m'',n'')}(x) \end{pmatrix} \times A^{(m',n')}_{\mu}(x), \notag \\
U_{R(m,n;m'',n'')}^{(m',n')} =
\begin{pmatrix} v_{(m,n;m'',n'')}^{(m',n')} & 0 \\ 0 & V_{(m,n;m'',n'')}^{(m',n')} \end{pmatrix},
\label{eq:general-amu-vertex}
\end{align}
where $g_4 \equiv \frac{g}{2 \pi R}$ is the four-dimensional gauge coupling and the vertex functions are given as
\footnote{{Here we use the abbreviation for Kronecker's delta : $\delta_m \equiv \delta_{m,0}$.}}
\begin{align}
V_{(m,n;m'',n'')}^{(m',n')} &=
\delta_{m+m'-m''} \delta_{n+n'-n''} + \delta_{m+n'-m''} \delta_{n-m'-n''} \notag \\
&+ \delta_{m+n'-n''} \delta_{n-m'+m''} + \delta_{m-m'+m''} \delta_{n-n'+n''} + \delta_{m-m'+n''} \delta_{n-n'-m''} \notag \\
&+ \delta_{m-m'-m''} \delta_{n-n'-n''} + \delta_{m-m'-n''} \delta_{n-n'+m''} + \delta_{m-n'+n''} \delta_{n+m'-m''} \notag \\
&+ \delta_{m-n'-m''} \delta_{n+m'-n''}, \label{eq:deltas-1}
\end{align}
\begin{align}
v_{(m,n;m'',n'')}^{(m',n')} &=
\delta_{m+m'-m''} \delta_{n+n'-n''} + \delta_{m+n'-m''} \delta_{n-m'-n''} \notag \\
&+i \delta_{m+n'-n''} \delta_{n-m'+m''} - \delta_{m-m'+m''} \delta_{n-n'+n''} -i \delta_{m-m'+n''} \delta_{n-n'-m''} \notag \\
&+ \delta_{m-m'-m''} \delta_{n-n'-n''} +i \delta_{m-m'-n''} \delta_{n-n'+m''} -i \delta_{m-n'+n''} \delta_{n+m'-m''} \notag \\
&+ \delta_{m-n'-m''} \delta_{n+m'-n''} \label{eq:deltas-2}.
\end{align}
The interaction vertices including at least one zero-mode are also given by the same form as (\ref{eq:general-amu-vertex}), but 
the vertex functions are different by factor 2 from what we obtain by formally generalizing (\ref{eq:deltas-1}) and (\ref{eq:deltas-2}) to the case of zero-mode: 
\begin{align}
&V_{(m,n;m'',n'')}^{(0,0)} = 2 \delta_{m,m''} \delta_{n,n''} \quad (m,m'',n,n'' \geq 0), \notag \\
&v_{(m,n;m'',n'')}^{(0,0)} = 2 \delta_{m,m''} \delta_{n,n''} \quad (m,m'' \geq 1\ ,\ n,n'' \geq 0), \notag \\
&V_{(0,0;m'',n'')}^{(m',n')} = 2 \delta_{m',m''} \delta_{n',n''} \quad (m' \geq 1\ ,\ m'' \geq 0\ ,\ n',n'' \geq 0), \notag \\
&V_{(m,n;0,0)}^{(m',n')} = 2 \delta_{m,m'} \delta_{n,n'} \quad (m \geq 0\ ,\ m' \geq 1\ ,\ n,n' \geq 0).
\label{eq:zerovertex}
\end{align}
In the case of the left-handed current, the $A_{{\mu}}$ interaction is written as
\begin{align}
&{\sum_{m,m',m''=1}^{\infty} \sum_{n,n',n''=0}^{\infty}}
\frac{g_4}{2} \begin{pmatrix} \overline{\psi^{'(m,n)}}(x) & \overline{\Psi^{'(m,n)}}(x) \end{pmatrix}
U_{L(m,n;m'',n'')}^{(m',n')} \gamma^{\mu} L
\begin{pmatrix} \psi^{'(m'',n'')}(x) \\ \Psi^{'(m'',n'')}(x) \end{pmatrix} \times A_{\mu}^{(m',n')}(x), \notag \\
& \quad U_{L(m,n;m'',n'')}^{(m',n')} = \hat{U}^{(m,n) \dagger} U_{R(m,n;m'',n'')}^{(m',n')} \hat{U}^{(m'',n'')}.
\label{lefthandedcurrent}
\end{align}
Actually, in a process where only the left-handed current appears without any chirality flip, the unitary matrices $\hat{U}$
 are irrelevant, 
\beq
\cdots \hat{U}^{(m,n)} \hat{U}^{(m,n) \dagger} U_{R(m,n;m'',n'')}^{(m',n')} \hat{U}^{(m'',n'')} \hat{U}^{(m'',n'') \dagger} \cdots = \cdots U_{R(m,n;m'',n'')}^{(m',n')} \cdots.
\eeq
This reflects the fact that the freedom of $V^{(m,n)}$ in (\ref{eq:left-rotation}) enables us to choose $V^{(m,n)} = \sigma_{3} U^{(m,n)\dagger}$ if we wish, so that $\hat{U}^{(m,n)}$ appears in the right-handed current instead of the left-handed current. On the other hand, in processes with chirality flip, the matrices $\hat{U}^{(m,n)}$ can not be eliminated and  describe the amplitudes.

From (\ref{eq:u-matrix}), (\ref{eq:general-amu-vertex}), (\ref{eq:zerovertex}) and (\ref{lefthandedcurrent}), we find that the interaction vertex of the ordinary photon $A_{\mu}^{(0)}$ takes the usual QED form:
\beq 
g_{4} \left(\sum_{m=0}^{\infty} \sum_{n=0}^{\infty} \overline{\Psi'^{(m,n)}} \gamma^{\mu} \Psi'^{(m,n)} + \sum_{m=1}^{\infty} \sum_{n=0}^{\infty} \overline{\psi'^{(m,n)}} \gamma^{\mu} \psi'^{(m,n)} \right) A_{\mu}^{(0)}. 
\eeq

Next, we discuss the interaction vertices of $A_{\omega}$ and $A_{\overline{\omega}}$.
The relevant part of the lagrangian is
\beq
\sqrt{2} g \left\{ i \overline{\psi} \tilde{\Psi} A_{\omega} + \mathrm{h.c.} \right\}.
\eeq
Here $A_{\omega}$ is mode expanded in terms of $G^{(m,n)}$ and $H^{(m,n)}$ as
\beq
A_{\omega}(x,y,z) = \frac{1}{2 \pi R} \sum_{m=1}^{\infty} \sum_{n=0}^{\infty} \frac{m-in}{\sqrt{2(m^2+n^2)}}
\left\{ G^{(m,n)}(x) -i H^{(m,n)}(x) \right\}
\times \left[ \sin\left(\frac{my+nz}{R}\right) +i \sin\left(\frac{ny-mz}{R}\right) \right].
\eeq
We thus realize that once we get the vertex function for $G^{(m',n')}(x)$, then the vertex function for $H^{(m',n')}(x)$ is readily obtained by multiplying by $-i$.

As $A_{\omega}$ has $Z_4$-eigenvalue $-i$, the vertex functions can be written in terms of $v_{(m,n;m',n')}^{(m'',n'')}$,
obtained by an exchange of $(m',n') \leftrightarrow (m'',n'')$ in (\ref{eq:deltas-2}):
\beq
{\sum_{m,m',m''=1}^{\infty} \sum_{n,n',n''=0}^{\infty}}
\frac{g_4}{2} \frac{m'-in'}{\sqrt{m'^2 + n'^2}} \overline{\psi^{(m,n)}}(x)  i v_{(m,n;m',n')}^{(m'',n'')} \tilde{\Psi}^{(m'',n'')}(x)
\left\{ G^{(m',n')}(x) -i H^{(m',n')}(x) \right\} + \mathrm{h.c.}.
\eeq
Rewriting in terms of mass eigenstates for the fermions, we get (for non-zero KK modes)

\begin{align}
&{\sum_{m,m',m''=1}^{\infty} \sum_{n,n',n''=0}^{\infty}}
\frac{g_4}{2} \frac{1}{\sqrt{m'^2+n'^2}} \Bigg\{
\begin{pmatrix} \overline{\psi^{'(m,n)}}(x) & \overline{\Psi^{'(m,n)}}(x) \end{pmatrix} \notag \\
&\times \begin{pmatrix} 
- \frac{m'-in'}{m_f^{(m'',n'')}} \frac{m''+in''}{R}  {i} v_{(m,n;m',n')}^{(m'',n'')} & - \frac{m'-in'}{m_f^{(m'',n'')}} m_B  {i} v_{(m,n;m',n')}^{(m'',n'')} \\
+ \frac{m'+in'}{m_f^{(m'',n'')}} m_B ({-i} )v_{(m'',n'';m',n')}^{(m,n)\ast} & - \frac{m' +i n'}{m_f^{(m'',n'')}} \frac{m''-in''}{R}
({-i} )v_{(m'',n'';m',n')}^{(m,n)\ast}
\end{pmatrix} L
\begin{pmatrix} \psi^{'(m'',n'')}(x) \\ \Psi^{'(m'',n'')}(x) \end{pmatrix} \times G^{(m',n')}(x) \notag \\
&  -i \begin{pmatrix} \overline{\psi^{'(m,n)}}(x) & \overline{\Psi^{'(m,n)}}(x) \end{pmatrix} \notag \\
&\times \begin{pmatrix} 
- \frac{m'-in'}{m_f^{(m'',n'')}} \frac{m''+in''}{R} {i} v_{(m,n;m',n')}^{(m'',n'')} & - \frac{m'-in'}{m_f^{(m'',n'')}} m_B {i} v_{(m,n;m',n')}^{(m'',n'')} \\
- \frac{m'+in'}{m_f^{(m'',n'')}} m_B ({-i} )v_{(m'',n'';m',n')}^{(m,n)\ast} & + \frac{m' +i n'}{m_f^{(m'',n'')}} \frac{m''-in''}{R}
({-i} )v_{(m'',n'';m',n')}^{(m,n)\ast}
\end{pmatrix} L
\begin{pmatrix} \psi^{'(m'',n'')}(x) \\ \Psi^{'(m'',n'')}(x) \end{pmatrix} \times H^{(m',n')}(x) \Bigg\}  \notag 
\end{align}
\begin{align}
+&{\sum_{m,m',m''=1}^{\infty} \sum_{n,n',n''=0}^{\infty}}
\frac{g_4}{2} \frac{1}{\sqrt{m'^2+n'^2}} \Bigg\{
\begin{pmatrix} \overline{\psi^{'(m,n)}}(x) & \overline{\Psi^{'(m,n)}}(x) \end{pmatrix} \notag \\
&\times \begin{pmatrix} 
- \frac{m'+in'}{m_f^{(m,n)}} \frac{m-in}{R} ({-i} )v_{(m'',n'';m',n')}^{(m,n)\ast} & + \frac{m'-in'}{m_f^{(m,n)}} m_B {i}  v_{(m,n;m',n')}^{(m'',n'')} \\
- \frac{m'+in'}{m_f^{(m,n)}} m_B ({-i} ) v_{(m'',n'';m',n')}^{(m,n)\ast} & - \frac{m' -i n'}{m_f^{(m,n)}} \frac{m+in}{R}
{i}  v_{(m,n;m',n')}^{(m'',n'')}
\end{pmatrix} R
\begin{pmatrix} \psi^{'(m'',n'')}(x) \\ \Psi^{'(m'',n'')}(x) \end{pmatrix} \times G^{(m',n')}(x) \notag \\
&  -i \begin{pmatrix} \overline{\psi^{'(m,n)}}(x) & \overline{\Psi^{'(m,n)}}(x) \end{pmatrix} \notag \\
&\times \begin{pmatrix} 
+ \frac{m'+in'}{m_f^{(m,n)}} \frac{m-in}{R} ({-i} )v_{(m'',n'';m',n')}^{(m,n)\ast} & + \frac{m'-in'}{m_f^{(m,n)}} m_B {i} v_{(m,n;m',n')}^{(m'',n'')} \\
+ \frac{m'+in'}{m_f^{(m,n)}} m_B ({-i} )v_{(m'',n'';m',n')}^{(m,n)\ast} & - \frac{m' -i n'}{m_f^{(m,n)}} \frac{m+in}{R}
{i}  v_{(m,n;m',n')}^{(m'',n'')}
\end{pmatrix} R
\begin{pmatrix} \psi^{'(m'',n'')}(x) \\ \Psi^{'(m'',n'')}(x) \end{pmatrix} \times H^{(m',n')}(x) \Bigg\}. 
\label{scalarvertices}
\end{align}
The interaction vertices including at least one zero-mode are also given by the same form as (\ref{scalarvertices}), by use of (\ref{eq:zerovertex}).  

\begin{figure}[H]
 \begin{minipage}{0.5\hsize}
  \begin{center}
   \includegraphics[width=80mm]{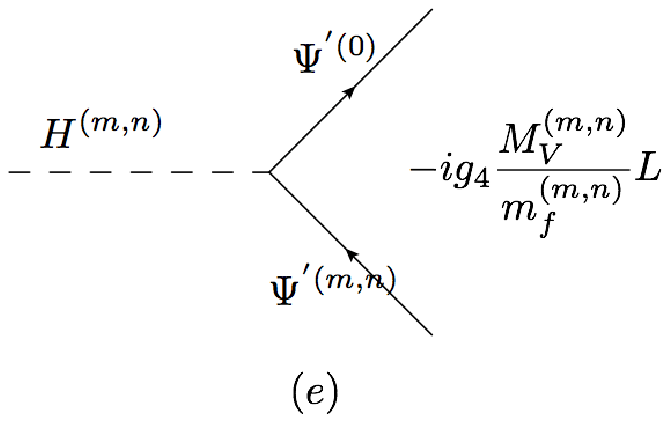}
  \end{center}
 \end{minipage}
 \begin{minipage}{0.5\hsize}
  \begin{center}
   \includegraphics[width=100mm]{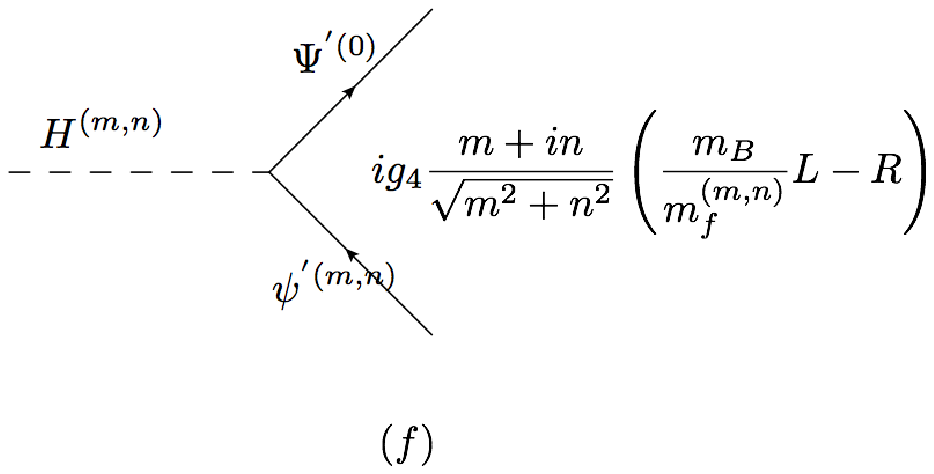}
  \end{center}
 \end{minipage}
\caption{The interaction vertices where one of the external fermion lines is the zero-mode $\Psi'^{(0)}$.} 
\label{fig:feynman-rules}
\end{figure}

It is interesting to note that a sort of ``equivalence theorem'' holds concerning the interaction vertices of non-zero KK modes $A_{\mu}^{(m,n)}$ and $G^{(m,n)}$, which are expected to hold as the result of Higgs-like mechanism operative in the sector of massive gauge-Higgs bosons. For illustrative purpose, we focus on the interaction vertices where one of external fermion lines is the zero-mode $\Psi'^{(0)}$ shown in Fig.{{\ref{fig:feynman-rules}}}, which are easily obtained from Eq.s (\ref{eq:u-matrix}), (\ref{eq:general-amu-vertex}), 
(\ref{eq:zerovertex}), (\ref{lefthandedcurrent}) and (\ref{scalarvertices}).    

\begin{figure}[H]
 \begin{minipage}{0.5\hsize}
  \begin{center}
   \includegraphics[width=85mm]{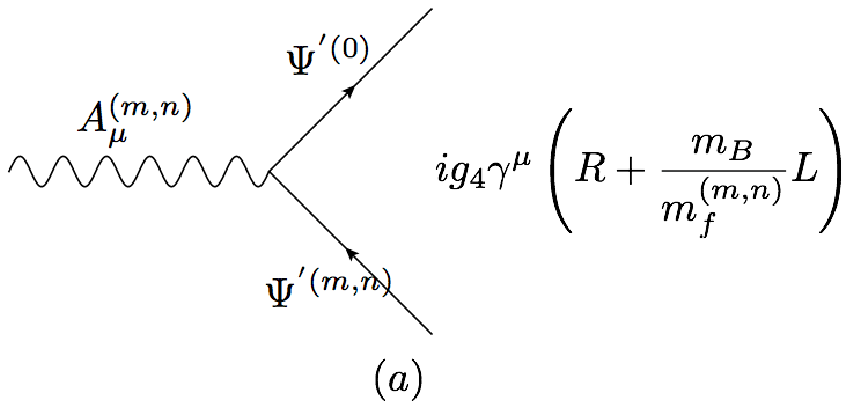}
  \end{center}
 \end{minipage}
 \begin{minipage}{0.5\hsize}
  \begin{center}
   \includegraphics[width=85mm]{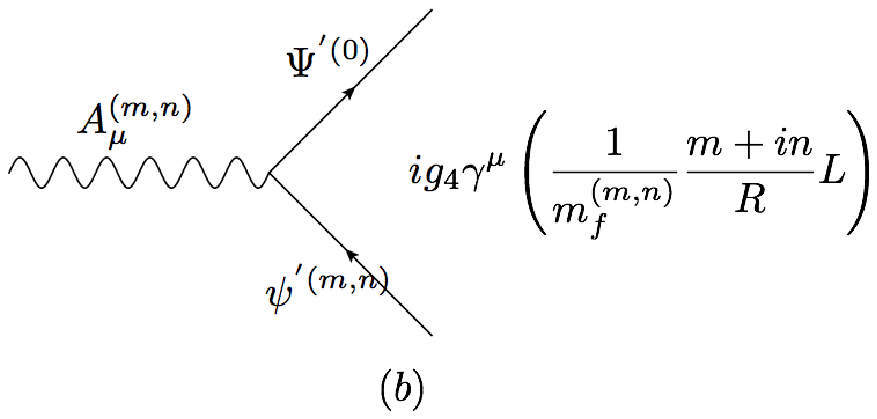}
  \end{center}
 \end{minipage}
\end{figure}

\begin{figure}[H]
 \begin{minipage}{0.5\hsize}
  \begin{center}
   \includegraphics[width=80mm]{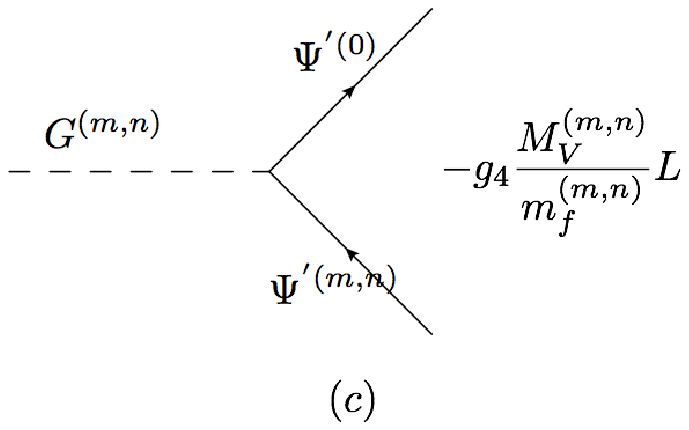}
  \end{center}
 \end{minipage}
 \begin{minipage}{0.5\hsize}
  \begin{center}
   \includegraphics[width=100mm]{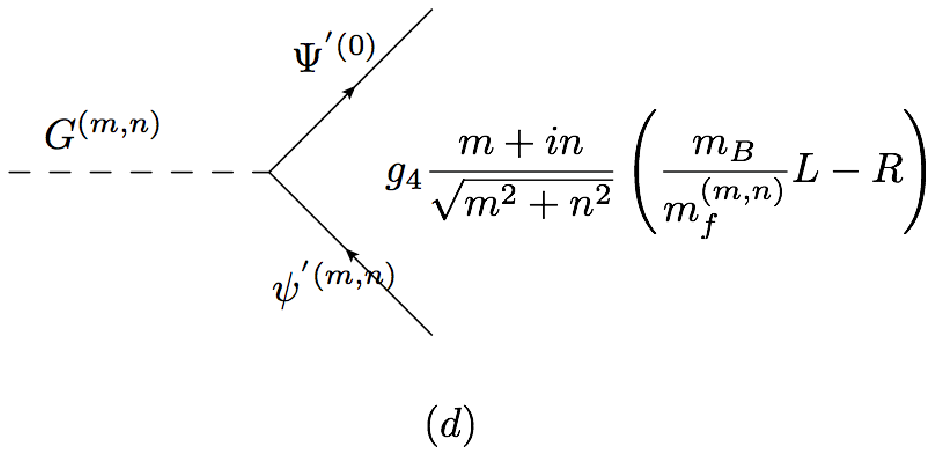}
  \end{center}
 \end{minipage}
\end{figure}

For instance, multiplying $\frac{k^{\mu}}{M_{V}^{(m,n)}}$ ($k^{\mu}$: the 4-momentum of $A_{\mu}^{(m,n)}$) to the current coupled with the massive photon, we get a relation by use of equations of motions for fermions
\beq 
\frac{k^{\mu}}{M_{V}^{(m,n)}} g_{4} \overline{\Psi^{(0)}} \gamma_{\mu} \left(R + \frac{m_{B}}{m_{f}^{(m,n)}} L \right) \Psi'^{(m,n)} = -g_{4} \frac{M_{V}^{(m,n)}}{m_{f}^{(m,n)}} \overline{\Psi^{(0)}} L \Psi'^{(m,n)},  
\eeq
where the r.h.s. just coincides with ($i$ times) the current coupled with $G^{(m,n)}$.  

\section{CP-violating photon interaction}

Now we are ready to confirm that CP violation is realized in our theory by showing that the imaginary couplings
in the interaction vertices remain even after the re-phasing of fermions {\cite{KM}}.

As we have seen in the previous section, the interaction vertices of KK modes of $A_{\mu} , G$ and $H$ are
rather complicated. We thus restrict ourselves to the interaction vertices of $A_{\mu}$, although the Yukawa couplings of KK modes of $G$ and $H$ also violate CP, as is suggested by the equivalence theorem.  
To make the analysis as transparent as possible, we consider only the right-handed current of $\psi'$
coupled to $A_{\mu}$, as shown in (\ref{eq:general-amu-vertex}), since the corresponding current due to $\Psi'$
described by $V$ has no phases. If all of $n,n'$ and $n''$ are set to $0$ the interaction vertex becomes real, as is 
seen in (\ref{eq:deltas-2}) and (\ref{eq:zerovertex}).
Thus we consider the case where $n=n'=0$ but $n'' \geq 1$, as the simplest non-trivial possibility:
\beq
\sum_{m,m',m'',n''=1}^{\infty} \frac{g_4}{2} \overline{\psi^{'(m)}} v_{(m;m'',n'')}^{(m')} \gamma^{\mu} R \psi^{'(m'',n'')} A_{\mu}^{(m')},
\label{eq:re-phasing-vertex}
\eeq
\beq
v_{(m;m'',n'')}^{(m')} \equiv v_{(m,0;m'',n'')}^{(m',0)} = i \delta_{m,n''} \delta_{m',m''} + \delta_{m,m''} \delta_{m',n''}, \label{eq:re-phasing-v}
\eeq
where we use the notation $\psi^{'(m)} \equiv \psi^{'(m,0)}$, etc.
Our task is to see whether the phase $i$ in (\ref{eq:re-phasing-v}) can be eliminated by suitable re-phasing of $\psi^{'(m)}$ and $\psi^{'(m'',n'')}$, or not.
More explicitly (\ref{eq:re-phasing-vertex}) is written, by use of (\ref{eq:re-phasing-v}), as
\beq
\sum_{m,m'=1}^{\infty} \frac{g_4}{2} \left[ \overline{\psi^{'(m)}} \gamma^{\mu} R \psi^{'(m,m')} A_{\mu}^{(m')} + i \overline{\psi^{'(m')}} \gamma^{\mu} R \psi^{'(m,m')} A_{\mu}^{(m)}  \right]. 
\label{singlesum} 
\eeq 
Write the re-phasing by use of phases $\phi_{m}$ and $\phi_{m,m'}$ as
\beq
\psi^{'(m)} \rightarrow e^{i \phi_m} \psi^{'(m)}\ ,\ \psi^{'(m,m')} \rightarrow e^{i \phi_{m,m'}} \psi^{'(m,m')}.
\eeq
For the case of $m=m'$, two interaction terms in (\ref{singlesum}) are actually identical and the resulting complex coupling $(1+i)g_4$ can be made real by re-phasing satisfying a condition
\beq
\phi_m - \phi_{m,m} = \frac{\pi}{4} \quad (\text{mod}\  \pi), 
\label{eq:re-phasing-cond1}
\eeq
where $\text{mod}\  \pi$ reflects the freedom to add an arbitrary multiple of $\pi$.  
For $m \not= m'$, the two terms in (\ref{singlesum}) are mutually independent, and we get two independent
conditions in order to eliminate the CP phases from the interaction lagrangian:
\beq
\phi_m - \phi_{m,m'} = 0 \quad (\text{mod}\ \pi),
\label{eq:re-phasing-cond2}
\eeq
\beq
\phi_{m'} - \phi_{m,m'} = \frac{\pi}{2} \quad (\text{mod}\ \pi).
\label{eq:re-phasing-cond3}
\eeq
The condition (\ref{eq:re-phasing-cond1}) can be trivially satisfied. Namely, for given $\phi_m$,
we can always find the solution of $\phi_{m,m}$. The combination of (\ref{eq:re-phasing-cond2}) and (\ref{eq:re-phasing-cond3}), however, give rise to non-trivial conditions for $\phi_m$:
\beq
\phi_{m} - \phi_{m'} = \frac{\pi}{2} \quad (\text{mod}\ \pi).
\label{eq:re-phasing-result1}
\eeq
Since this condition should be met for arbitrary $m$ and $m'$ $(m \not= m')$, we realize that all $\phi_m$ $(m \not= m')$ must be the same $(\text{mod}\ \pi)$ for given $m'$. As the $m'$, in turn, can be arbitrary we find that all $\phi_m$
should be the same $(\text{mod}\ \pi)$:
\beq
\phi_1 = \phi_2 = \cdots \quad (\text{mod}\ \pi).
\label{eq:re-phasing-result2}
\eeq
On the other hand (\ref{eq:re-phasing-result2}) clearly contradicts (\ref{eq:re-phasing-result1}).
Thus we conclude that the conditions (\ref{eq:re-phasing-cond2}) and (\ref{eq:re-phasing-cond3}) are
incompatible with each other and the CP-violating phases cannot be removed by the re-phasing. Let us note that 
if some part of the interaction lagrangian violates CP, so does the whole lagrangian. 
Hence we have confirmed that CP is violated in our model.

\section{Re-phasing invariant quantities}

Although we have shown that CP is violated in our model by considering a concrete example, for the completeness of the argument it would be desirable to identify re-phasing invariant CP violating parameters, $\grave{a}$ la the Jarlskog parameter in the Kobayashi-Maskawa model {\cite{Jar}}. It would be also helpful in understanding what are the physical invariants appearing in the amplitudes of CP violating processes.  
As a matter of fact, in our model the free lagrangian of fermions has a larger symmetry than
the re-phasing: it is invariant under the unitary transformation described by $V^{(m,n)}$ in (\ref{eq:left-rotation}). Thus the CP violating observables should be invariant under the transformation and therefore they are written as the imaginary parts of the trace of the products of the matrices $U_{R,L(m,n;m'',n'')}^{(m',n')}$ (for the case of $A_{\mu}$ interaction).

First, let us focus on the observables due to $A_{\mu}^{(m',n')}$ interactions where each fermion propagator possesses no chirality flip.
Let us note that as far as processes without chirality flip are considered there is no difference between the processes due to right- and left- handed currents, as we have already discussed. Thus, here we consider only the processes due to the right-handed current.

Our task is to find non-vanishing imaginary parts of the trace of the products of  $U_{R,L(m,n;m'',n'')}^{(m',n')}$,
appearing in the $A_{\mu}^{(m',n')}$ vertex (see (\ref{eq:general-amu-vertex}), (\ref{eq:deltas-1}) and (\ref{eq:deltas-2})):
\beq
\text{Im Tr} \left( U_{R(m,n;m'',n'')}^{(m',n')} U_{R(m'',n'';m'''',n'''')}^{(m''',n''')} \cdots \right)
= \text{Im} \left( v_{(m,n;m'',n'')}^{(m',n')} v_{(m'',n'';m'''',n'''')}^{(m''',n''')} \cdots \right),
\label{eq:Jarlskog1}
\eeq
\begin{figure}[H]
 \begin{center}
  \includegraphics[width=65mm]{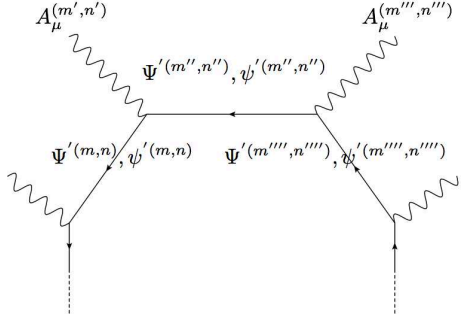}
 \end{center}
\caption{{An example of Feynman diagram describing re-phasing invariant quantities}.}
 \label{fig:J}
\end{figure}
\noindent
corresponding to the Feynman diagram shown in Fig.\ref{fig:J}.
In Fig.\ref{fig:J}, the fermions form a closed loop, thus making (\ref{eq:Jarlskog1}) invariant under the unitary transformation,
\beq
U_{R(m,n;m'',n'')}^{(m',n')} \rightarrow V^{(m,n)\dagger} U_{R(m,n;m'',n'')}^{(m',n')} V^{(m'',n'')},
\eeq
corresponding to the freedom of $V^{(m,n)}$ in (\ref{eq:left-rotation}).
In the case of Kobayashi-Maskawa model,
the Jarlskog parameter arises only at the 4-th order of the KM matrix elements $V^{KM}_{i \alpha}$;

\noindent
$J = |\text{Im}(V^{KM}_{i \alpha} V^{KM \ast}_{j \alpha} V^{KM}_{j \beta} V^{KM \ast}_{i \beta})|\ (i \not= j,\alpha \not= \beta)$.
In our model, however, the imaginary part is found to arise already at the second order of $v_{R(m,n;m'',n'')}^{(m',n')}$, 
\beq
\text{Im} \left( v_{(m,n;m'',n'')}^{(m',n')} v_{(m'',n'';m,n)}^{(m''',n''')}  \right), 
\eeq
because of the variety of the KK modes $A_{\mu}^{(m',n')}$. For instance if we set 
\beq
m=1 , n=a , m' = 2 , n' = a+2 , m''=1 , n'' = 2 , m''' = a+1 , n''' = 1,
\eeq
with an arbitrary positive integer $a$ $(\not = 2)$, we find
\beq
\text{Im} \left( v_{(1,a;1,2)}^{(2,a+2)} v_{(1,2;1,a)}^{(a+1,1)}  \right) = 1.
\eeq

\section{A brief comment on the EDM of electron}  

Arguments given above have shown that the CP violating phases remain even after the re-phasing of the fields and therefore CP is broken as the consequence of the compactification on $T^{2}/Z_{4}$.  As a concrete example of a CP violating observable 
here we comment on the electric dipole moment (EDM) of electron, since the EDM does not need the fermion generation structure, which
is ignored in our model. 

\begin{figure}[H]
 \begin{minipage}{0.33\hsize}
  \begin{center}
   \includegraphics[width=50mm]{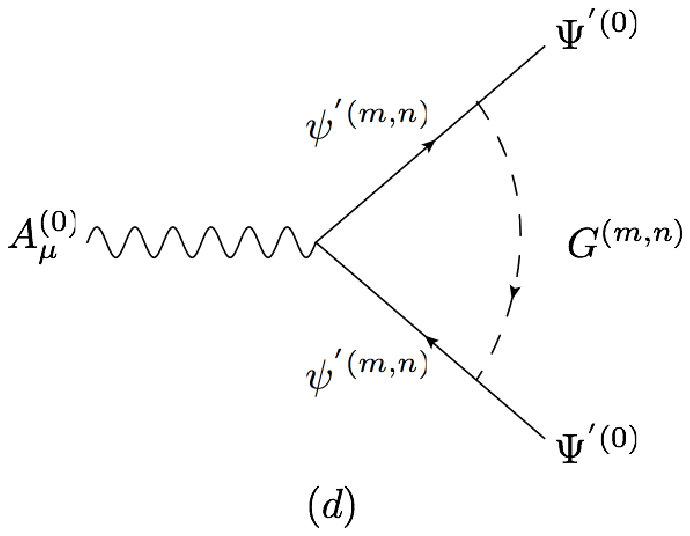}
  \end{center}
 \end{minipage}
 \begin{minipage}{0.33\hsize}
 \begin{center}
  \includegraphics[width=50mm]{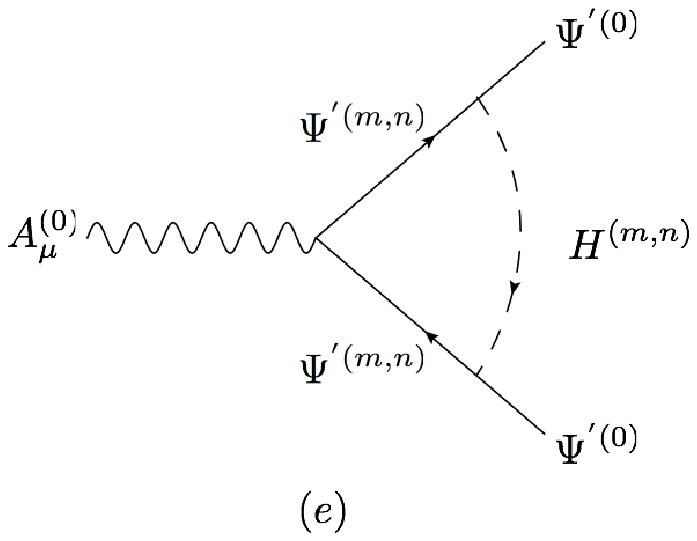}
 \end{center}
 \end{minipage}
 \begin{minipage}{0.33\hsize}
 \begin{center}
  \includegraphics[width=50mm]{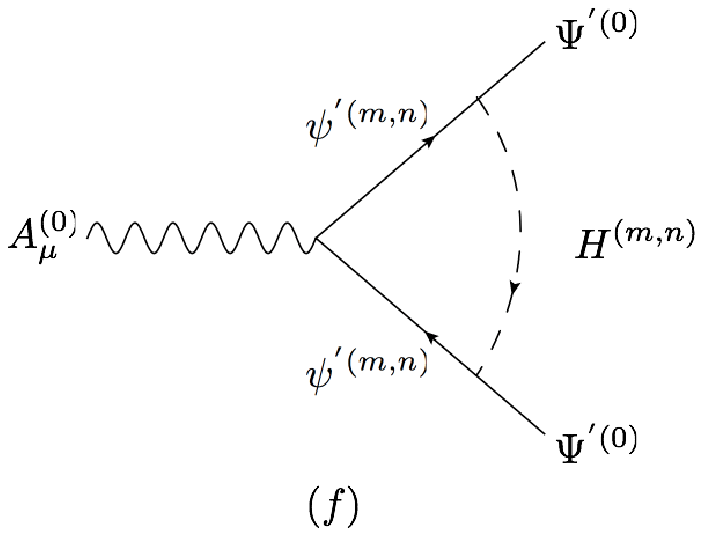}
 \end{center}
 \end{minipage}
\caption{Feynman diagrams for electron EDM at the 1-loop level.}
\label{fig:EDM1-loop}
\end{figure}

We first focus on the possible 1-loop contributions to the EDM, where the intermediate states are the non-zero KK modes of electron and gauge-Higgs bosons. The relevant 1-loop diagrams are those in Fig.{\ref{fig:EDM1-loop}}.  

\begin{figure}[H]
 \begin{minipage}{0.33\hsize}
  \begin{center}
   \includegraphics[width=50mm]{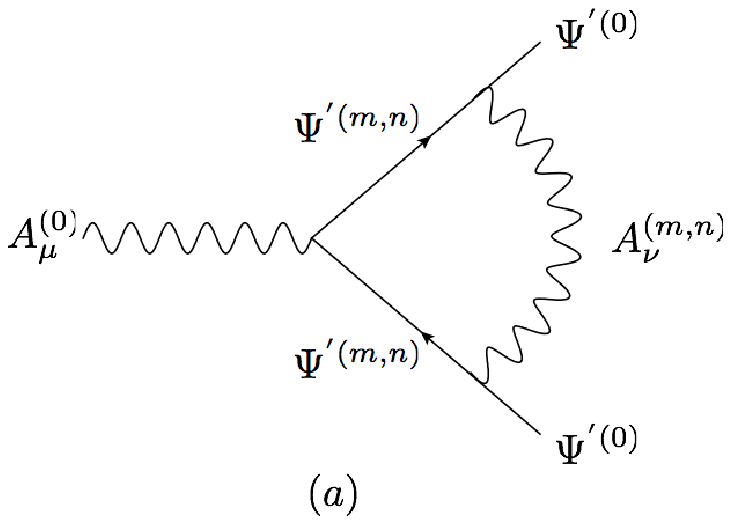}
  \end{center}
 \end{minipage}
 \begin{minipage}{0.33\hsize}
 \begin{center}
  \includegraphics[width=50mm]{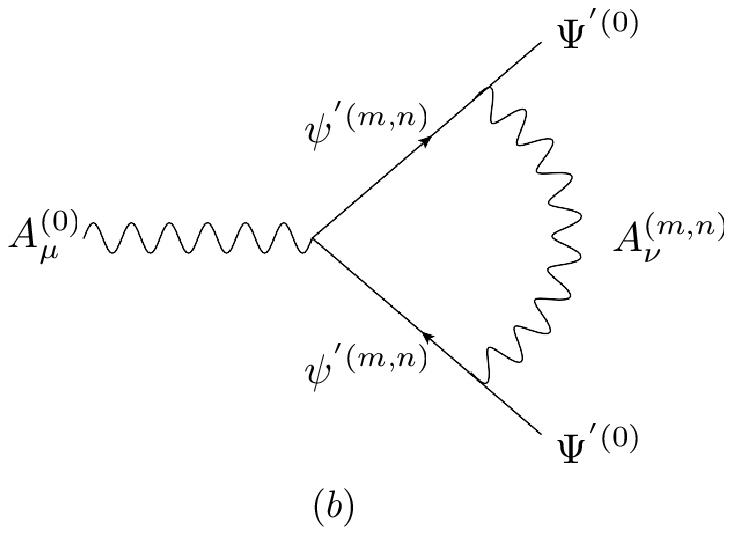}
 \end{center}
 \end{minipage}
 \begin{minipage}{0.33\hsize}
 \begin{center}
  \includegraphics[width=50mm]{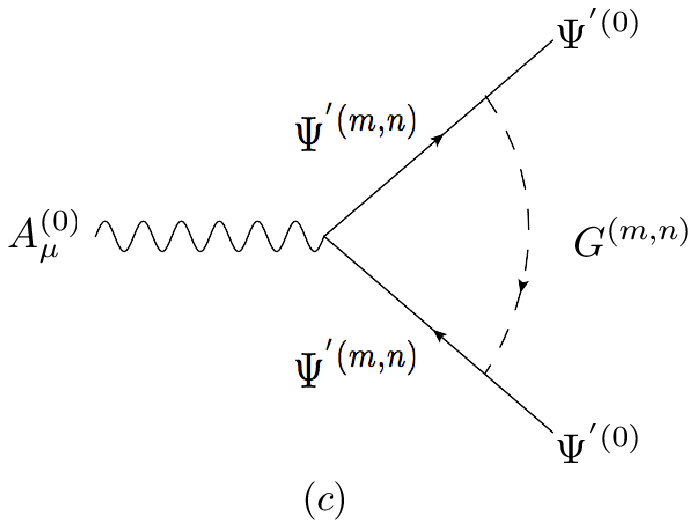}
 \end{center}
 \end{minipage}
\end{figure}

Since the Feynman diagrams are divided into two types, i.e. diagrams with the exchanges of four-dimensional vector and four-dimensional scalar, it may be useful to derive general formulas for the amplitudes of these two types of diagrams, due to generic interactions, $\overline{\tilde{\psi}} \gamma_{\mu}(aL + bR)\Psi'^{(0)}V^{\mu} + h.c.$ and $\overline{\tilde{\psi}} (a'L + b'R)\Psi'^{(0)} S + h.c.$, respectively, where $\tilde{\psi}, \ V^{\mu}, \ S$ denote intermediate states of fermion, four-dimensional vector and four-dimensional scalar. The general formulas for these two types of diagrams are known to be proportional to $2i \mbox{Im} (a b^{\ast})$ and $2i \mbox{Im} (a' b'^{\ast})$, respectively \cite{ALM4}.  

This observation immediately leads to an important conclusion that we do not get the EDM at least at the 1-loop level. 
Namely, all of $2i \mbox{Im} (a b^{\ast})$ and $2i \mbox{Im} (a' b'^{\ast})$ obtained from the interaction vertices shown in Fig.{\ref{fig:feynman-rules}} vanish. Let us remember that to get an EDM both $P$ and $CP$ symmetries should be broken. The origin we find for the cancellation of the 1-loop contribution arises from the fact that the orbifolding does not violate $P$ symmetry, while it does break $CP$, since as is seen in (\ref{CandCP}) the extra-space coordinates are invariant under the $P$ transformation. At the first glance this argument seems to contradict the fact that 
the right- and left-handed currents coupled to the non-zero KK modes of the photon are not identical (see (\ref{eq:general-amu-vertex}) and (\ref{lefthandedcurrent})). We, however, realize that if we choose $V^{(m,n)} = \sigma_{3}U^{(m,n)\dagger}$, instead of $V^{(m,n)} = U^{(m,n)\dagger}$ in (\ref{eq:left-rotation}), the roles of right- and 
left-handed fermions are exchanged, compared with the case of (\ref{eq:u-hat-basis}). This implies that the contribution to the EDM should be invariant under the exchange $a \leftrightarrow b$ and therefore $\mbox{Im} (a b^{\ast}) = \mbox{Im} (b a^{\ast}) = 0$. 

As far as the vanishing contribution to the EDM has its origin in the $P$ symmetry of the model, we anticipate that the EDM will not emerge even at the higher loop Feynman diagrams, though the explicit computations are desirable to settle the issue. 
Nevertheless, we still expect that EDM gets non-vanishing contribution as long as the $CP$ symmetry is violated by the orbifolding, once the model is made realistic in a way it can incorporate the standard model, where $P$ symmetry is broken.         

\section{Summary}  

In this paper we addressed the question of how CP violation is realized in the scenario of gauge-Higgs unification, where the interaction of the Higgs is governed by a gauge principle and therefore to get CP violating phases is a challenging issue. 

As a simple and non-trivial example we examined a 6-dimensional U(1) model compactified on an 2-dimensional orbifold $T^{2}/Z_{4}$. First we extended an argument of how four-dimensional CP transformation is related to the complex structure of the extra space and showed that the adopted $Z_{4}$ orbifolding is incompatible with such defined CP symmetry and therefore leads to CP violation. Next, we confirmed the expectation by extensively studying the 
interaction vertices derived from the overlap integrals over the extra-space coordinates of mode functions. We could get CP violating phases which do not vanish even after the possible re-phasing of the relevant fields. For completeness, we derived a re-phasing invariant CP violating parameter, following a similar argument in the Kobayashi-Maskawa model which led to the Jarlskog parameter.   

As a typical example of CP violating observable we made a brief comment on the EDM of electron in our model. It turned out that at the 1-loop level, the EDM gets no contributions. The origin of the vanishing EDM in our model was argued to be the fact that the orbifolding does not break the $P$ symmetry, while both of $P$ and $CP$ symmetries should be broken to get a non-vanishing EDM.  Nevertheless, the EDM is expected to get a non-vanishing contribution as long as the $CP$ symmetry is violated by the orbifolding, once the model is made realistic in a way that incorporates the standard model where $P$ symmetry is broken.         
The chiral theory with P violation will be realized, once we start from six-dimensional Weyl fermion with definite eigenvalue of $\Gamma_{7}$, instead of six-dimensional Dirac fermion, though in that case we have to ensure the cancellation of six-dimensional anomaly \cite{Dobrescu} by suitably choosing the matter content. 

An interesting candidate of such realistic higher dimensional gauge theory of the type 
discussed in this paper may be the theory based on the ``gauge-Higgs unification'' scenario. The scenario was proposed long time ago 
\cite{Manton, Fairlie, Hosotani,Antoniadis}, where the Higgs field is identified with the zero-mode of an extra spatial component of higher dimensional gauge fields. 
It has been revived as one of the attractive scenarios solving the hierarchy problem without invoking supersymmetry \cite{HIL}. This is based on the observation that the quantum correction to the Higgs mass is finite and insensitive to the ultra-violet (UV) cutoff of the theory thanks to the higher dimensional local gauge symmetry, in spite of the fact that higher dimensional gauge theories are generally regarded as non-renormalizable. Since then, many interesting works based on this scenario have appeared in the literature from various points of view \cite{KLY}-\cite{HSY}.

Strictly speaking, the U(1) model discussed in this paper is not a model of gauge-Higgs unification, as the extra-space component of gauge field does not have a zero-mode that behaves as a Higgs field. We, however, believe that the discussions of the mechanism of CP violation extended in this paper holds in general for the models of gauge-Higgs unification with larger gauge symmetries including that of the standard model, since the mechanism is based on the manner of compactification and does not depend on the choice of the gauge group. It, however, should be pointed out that the introduction of brane-localized fields and their interactions with bulk fields may be needed to make the theory realistic \cite{ACP}. The localized mixing mass parameter may 
become another source of CP violation, independent of the mechanism of CP violation due to the compactification discussed in this paper. 

It is interesting to note that the proposed mechanism of CP violation due to the $Z_{4}$ orbifolding does not need flavor or generation structure, as our U(1) model incorporates only 1 generation, i.e. the electron. The CP violation is achieved through the interactions including non-zero KK modes. From such a point of view, our mechanism of CP violation is quite different from that in the Kobayashi-Maskawa model. It will be an interesting and important question how the mechanism of CP violation can be extended when we include multiple generations.  Once the generations are introduced we will be able to discuss other well-known CP violating 
observables caused by flavor changing neutral current processes, such as $\epsilon$ in the neutral kaon system or CP asymmetries in B meson decays.     

\vspace{5mm}
\noindent{\large \bf Acknowledgments}
\vspace{3.5mm}

\noindent
We would like to thank W.J. Marciano for careful reading of the manuscript and for very useful comments. The work of the authors was supported in part by the Grant-in-Aid for Scientific Research
of the Ministry of Education, Science and Culture, No.18204024 and No.20025005.

\end{document}